\documentclass[%
preprint,
 amsmath,amssymb,
]{revtex4-1}

\usepackage{graphicx}
\usepackage{dcolumn}
\usepackage{bm}
\begin{document}


\title{High-dimensional measurement-device-independent quantum key distribution based on spatial basis}

\author{Wei Li$^{1,3}$}
 \email{alfred$_$wl@njupt.edu.cn}
 \author{Shengmei Zhao$^{1,2}$}%
 \email{zhaosm@njupt.edu.cn}
 \altaffiliation{Nanjing University of Posts and Telecommunications, Institute of Signal Processing and Transmission, Nanjing, 210003, China.}
\affiliation{Nanjing University of Posts and Telecommunications, Institute of Signal Processing and Transmission, Nanjing, 210003, China.}%
\affiliation{Nanjing University of Posts and Telecommunications, Key Lab Broadband Wireless Communication and Sensor Network, Ministy of Education, Nanjing, 210003, China.}%
\affiliation{Sunwave Communications Co., Hangzhou, 310053, China.}%

\date{\today}

\begin{abstract}
Improving the secret key rate is one of the vital issues in practical applications of quantum key distribution (QKD). In this paper, we propose an experimental scheme of high-dimensional measurement-device-independent quantum key distribution (MDI-QKD) based on spatial basis aiming to increase the key rate. Two groups of discrete position basis and momentum basis are applied as conjugate spaces to generate quantum secret key. The position states are transmitted by single-mode fibers and  represented by the index of the fiber, while the momentum state is a coherent superposition of the position states characterized by the phase gradient. The measurement of each momentum basis is realized by multi-slit diffraction. This experimental proposal can be implemented with standard optical elements. In addition to an enhanced key rate in a higher dimension, this high-dimensional MDI-QKD exhibits a comparable security level/performance compared to conventional polarization-based MDI-QKD.
\end{abstract}

\pacs{Valid PACS appear here}
\maketitle


\section{\label{sec:level1}Introduction}

\par Quantum key distribution (QKD) allows two far separated parties, say Alice and Bob, to share a string of secret random number in the presence of an eavesdropper. This random number string, known only to Alice and Bob, can be used to encrypt the message transmitted between them. In the BB84 QKD protocol\cite{charles2014quantum}, Alice randomly chooses one of a conjugate spaces and prepares her single-photon state in one basis before sending it to Bob. At the receiving end, Bob randomly selects a basis to measure the received photon state and distills the random number string through public discussion. The absolute security of QKD is guaranteed by the principle of quantum mechanics in the ideal case of single photon source and perfect single-photon detector. However, there is always a gap between theory and practical physical implementation, namely a weak coherent pulse is used to approximate the single-photon source, the detection device is imperfect. This provides chances for Eve to implement photon-number-splitting (PNS) attack\cite{norbert2002quantum,scarani2004quantum}, or detection side channel attacks\cite{fung2007phase,feihu2010experimental,zhao2008quantum,lydersen2010hacking,gerhardt2011full}. The recent proposed decoy state method combined with measurement-device-independent (MDI) QKD has been proved to have the potential to remove all the real-life loopholes \cite{wang2005beating,lo2005decoy,braunstein2012side,lo2012measurement,rubenook2013real,ferreira2013proof}. Meanwhile, enormous progress have been made on implementing MDI-QKD in experiments\cite{liu2013experimental,tang2014measurement,tang-2014-experimental,pirandola2015high,wang2015phase,comandar2016quantum,tang2016experimental}.

\par Even though the immunization of MDI-QKD to all detection attacks has been proved, but there is still an noticeable limitation that its secret key rate is much less than that of BB84 protocol. This shortage prevents it from the requirement of practical applications. Several attempts have been proposed to increase the key rate, which include parameter optimization\cite{ma2012statistical,xu2014protocol,yu2015statistical}, proposing asymmetric decoy-state MDI-QKD protocol\cite{yin2016measurement,zhou2016making,zhang2018biased}, increasing the system clock rate\cite{comandar2016quantum} and increasing the efficiency of the single-photon detector\cite{marsili2013detecting}. While the previously proposed MDI-QKD is based on the polarization of a photon state, an intrinsic property of light, which is a degree of freedom with a dimension of 2. Thus, a single photon could only carry at most one bit of information. Moreover, attempts to increase the information carrying ability have also been reported, such as the multiplexing of polarization and orbital angular momentum\cite{vallone2014free}, exploiting high dimensional degree of freedom\cite{chau2015quantum,chau2017experimentally,frederic2018experimental}. Compared to the polarization of light, one advantage of the extrinsic degree of freedoms is the unlimited high-dimensions, which could make a photon to carry high-bit information. As far as we know up to date, there has been no report on high-dimensional MDI-QKD, which both favor the security of quantum secret key sharing and high information transmission rate.

\par In this paper, we are going to present an experimental scheme of the high-dimensional MDI-QKD based on the spatial basis as an approach to increase the secret key rate. Position basis and momentum basis are used as $Z$ basis and $X$ basis, and to form conjugate spaces to implement secret key distribution. These two bases are connected by a discrete Fourier transformation. The measurement of the momentum state is realized by multi-slit diffraction, and the momentum basis is characterized by their propagation direction. The main drawback of this scheme is the finite overlap between adjacent diffraction peaks in the momentum measurement. However this drawback can be overcome by choosing the momentum basis sparsely for a larger dimension. Furthermore, we analyze the security of this high-dimensional MDI-QKD, and give a derivation of the secret key rate equation as well as a discussion about
the dependence of the secret key rate on the dimension and transmission distance.

\par Recalling that in the original BB84 QKD protocol, Alice should prepare a string of single photon state in two mutually unbiased basis (MUB). These MUB can be rectilinear basis ($Z$ basis) and the diagonal basis ($X$ basis) or the circular basis ($Y$ basis). While for high-dimensional degree of freedom, there are always more than 3 MUBs that can be applied for QKD\cite{durt2010on}. Here we choose discrete basis in position space and momentum space as the two MUBs. The discrete position space with a dimension of $N$ is chosen as the $Z$ basis, which is assigned as the computational basis, and the photon state in the j-th basis reads $\left | j \right \rangle_{Z}$, where $j$ ranges from 0 to $N-1$. The corresponding states in $X$ space are obtained by generalized discrete Fourier transformation, then we get the $k-th$ single-photon state in $X$ basis as:
\begin{equation}
\left | k \right \rangle_{X}=
\frac{1}{\sqrt{N}}\sum_{j=0}^{N-1}e^{i\frac{\pi\cdot 2k}{N}\cdot j}\left | j \right \rangle_{Z},
\end{equation}
From equation (1), it can be seen that the $X$ basis is a set of momentum spaces. In this paper, we apply the $X$ basis and $Z$ basis as the conjugate space to generate quantum secret key.

\begin{figure}[ht]
\centering
\includegraphics[width=120mm]{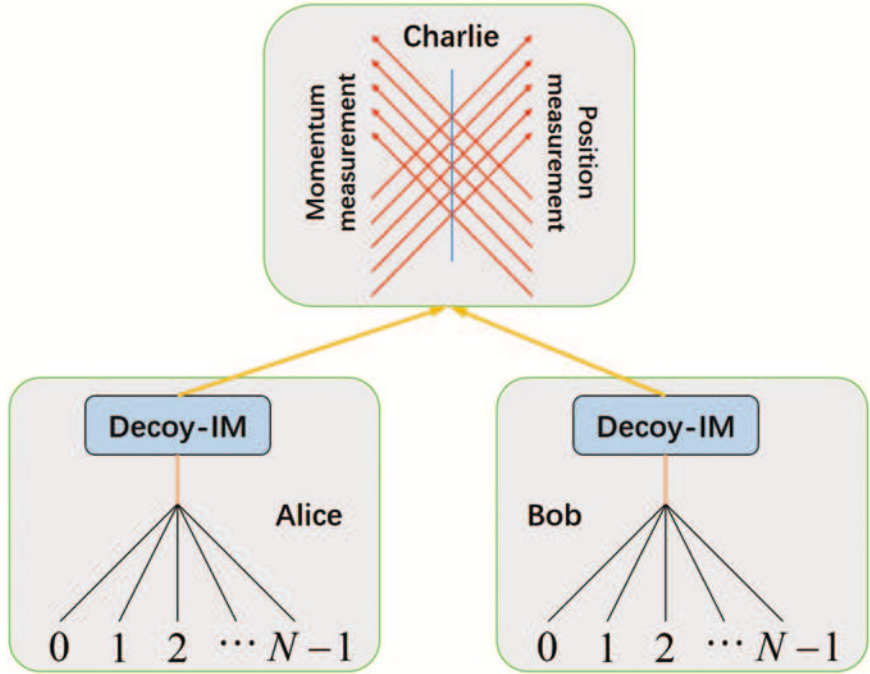}
\caption{Proposed experimental setup for high-dimensional position based MDI-QKD. Alice and Bob randomly prepare her weak-coherent pulse (WCP) in X basis or Z basis, then use intensity modulator (Decoy-IM) to generate decoy states before sending to the third party, Charlie, for joint basis measurement. Within the detection device, the signal states from Alice and Bob interfere at a multi-input multi-output beamsplitter which is followed by a momentum and a position measurement in two opposite sides. The momentum measurement is realized by multi-slits diffraction, while the position information is extracted by connecting the output ports of the fibers to a set of single-photon detectors. The measurement result at one instance is labeled as the indices of the registered photon detectors and the category of the measurement. A successful output corresponds the case that there are simultaneously two detectors fired in either the momentum measurement side or the position measurement side.}
\label{Fig.}
\end{figure}

\section{\label{sec:level1}Experimental Setup}

\par The proposed experiment setup for high-dimensional position based MDI-QKD is shown in Fig.1. The MDI-QKD scheme consists of three parties: Alice, Bob and Charlie. Alice and Bob share a string of secret random number, while Charlie plays the role of photon states measurement. In the position space (Z basis), the eigen-states are transmitted by different optical fibers, and in the momentum space (X basis), the basis is the coherent combination of the Z basis characterized by different phase gradients. Alice and Bob randomly prepare their phase randomized weak coherent pulses in either Z basis or X basis, and send them to the third party measurement agency, Charlie. The detection device of Charlie consists of a multi-input multi-output beam splitter, then followed by momentum and position measurements performed in two opposite sides. On the left side as shown in Fig.1, the momentum measurement is realized by multi-slits diffraction where different momentum states would be diffracted into different directions. On the other side, the position information is extracted by connecting the output ports of the fibers to a set of single-photon detectors. The measurement result is labeled as the indices of the single-photon detectors and the category of the measurement. For example a register in the $m$th detector in the position measurement side is labeled as $m_{Z}$, and a register in the $n$th detector in the momentum measurement side is labeled as $n_{X}$. An output is successful whenever any two detectors in either the position measurement side or the momentum measurement side are simultaneously registered.

\begin{figure}[ht]
\centering
\includegraphics[width=120mm]{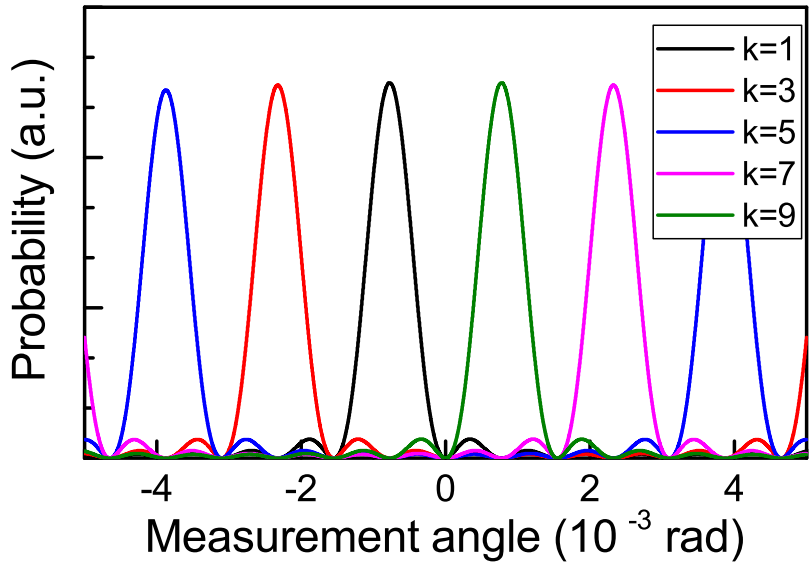}
\caption{Momentum basis measurement with a dimension of 10 through multi-slit diffraction. The single mode fiber for 1550 nm is used for the simulation. The width of each slit, i.e. the diameter of the inner core, is equal to 0.01 mm. The distance of adjacent slits is equal to the outer diameter (0.2 mm). }
\label{Fig.}
\end{figure}

\par When Charlie finishes the measurement, she broadcasts the results (the indices of the two registered detectors) to Alice and Bob through classical communication wires as well as the measurement basis that the results have been successfully registered. Then Alice and Bob announce in the public channel in what basis they have prepared their photon states. They keep the choices when their preparation are consistent with Charlie's measurement. According to Charlie's measurement result and the prepared basis, it is very likely that A and B will know each other's prepared photon state. The shared common private random number can be obtained by subtracting their prepared photon state from Charlie's measurement result. This step plays the same role as the Bell state analysis in the two-dimensional polarization based MDI-QKD\cite{lo2012measurement}. Note that the position measurement and momentum measurement here are spatially separated, which is different from polarization measurement by Bell state analysis. This difference lies in the fact that polarization is an intrinsic property of the photon state, and it has the same property as its conjugate space, like rectilinear basis and diagonal basis. However, for position basis and momentum basis which are extrinsic properties of a photon state with infinite dimensions, they behave accordingly.

\par Here we compare the security level of this high-dimensional MDI-QKD to the conventional two-dimensional polarization-based one. In the polarization based MDI-QKD, even though Eve knows exactly the Bell state of the photon sent by Alice and Bob, but his probability of acquiring the correct photon state from Alice and Bob is $50\%$. When the length of the random number string goes to infinity, then his probability of obtaining the correct shared random number string is zero. Similarly, in this high-dimensional MDI-QKD scheme, Eve may infer the product state of Alice and Bob, but he knows nothing at all about the exact state that belongs to Alice and Bob. As the final key is only extracted from one of them, Eve's probability of guessing right the photon state at each instance is also $50\%$. So the security level of our high-dimensional MDI-QKD is the same as the two-dimensional MDI-QKD based on polarization.

\par Fig.2 shows the simulation of momentum detection for single photon states with a dimension of $N=10$. To satisfy the spatial  indistinguishability between the photon states sent by Alice and Bob, all the photon states are all transmitted by single mode fibers. From this figure, it can be seen that with additional phase gradient, the diffraction peaks for momentum state $\left | k \right \rangle_{X}$ locate at an angle of $\frac{(Nj+k)\lambda }{lN}$, where $l$ is the distance between adjacent fibers, $j$ is an arbitrary integer, $N$ is the dimension of the basis, $\lambda$ is the wavelength of the photon state and $k$ is the index of the momentum mode.  In this simulation, we choose the width of single-mode fiber as 0.01 mm, the distance of adjacent fibers as 0.2 mm, the wavelength of the photon state as 1550 nm. The width of the central peak of the diffraction pattern is $\frac{2\lambda }{lN}$, which is twice the distance between the diffraction central peaks from adjacent momentum basis. Thus to reduce the overlap for momentum measurement, five out of ten momentum basis of $\left | k \right \rangle_{p}$ with $k=1,3,5,7,9$ were chosen. Besides, we have shown above that the security level dose not depend on the space dimension, thus such treatment is feasible.

\par Now we evaluate the performance of the high-dimensional MDI-QKD. Due to the different detection mechanism, we evaluate the data analysis in these conjugate bases separately. However, as position and momentum are continuous in nature, the diffraction pattern of the momentum measurement is periodical in detection angle, so the detection of single-photon state in momentum base is more complex and less efficient than that in the position space. Here we use the position basis as key generation basis, while the momentum basis is used for checking the existence of eavesdropping. In the entanglement purification protocol and quantum error correcting codes, there are two kinds of errors, bit error and phase error. The amount of information sacrificed for these processes should cover both of these error entropy. Once the key generation basis is selected, phase error corresponds to the bit error in its conjugate space. In the proof of the security of the BB84 QKD protocol, an amount of information equal to the error bit entropy in the conjugate space is sacrificed for privacy amplification. In this position based high-dimensional MDI-QKD, the phase error in position basis corresponds to the bit error in the momentum basis.

\par For both Z basis and X basis, an error corresponds to the case that the indices of the registered photon detectors are not in accordance with the photons' basis. Here we ignore the influence from dark count and stray light. In Z basis, as different basis states are transmitted in different fibers, there is no chance to have any crosstalk between them. In this case, no matter what the photon number states sent by Alice and Bob are, the response of the detector must be in accordance with the basis of the photon state. In X basis measurement, the output direction of the photon state strongly depends on the basis of each state. After the multi-slit diffraction, the photon states are spatially separated according to their propagation direction, so the multi-photon effect will not exert any influence on the measurement results. The only error for X basis measurement comes from the overlap between the diffraction peaks of different basis states. The amount of overlap is determined by the leakage of detection probability to its the secondary peaks. As shown in Fig.2, we choose the X basis with an interval of 2. If $N$ is large enough, the leakage of the photon state to the neighboring detector finally reduces to about $3\%$. It means that the total probability for error detection in momentum basis is about $6\%$. Furthermore, for a large enough dimension $N$,  the crosstalk in the momentum measurement can be reduced to almost 0 through choosing the checking X basis sparsely. For example, the X basis is chosen with an interval of $3$, then the number of basis in X space is one third of that in Z space, $N_{X}=\frac{1}{3}N_{Z}$.

\section{\label{sec:level1} key generation rate}

\par The photon states in Z basis have a higher key productiveness than in X basis, so we prefer to choose the Z basis as the key generation and the X basis used for eavesdropping checking. Assume Alice prepares her photon state in one of the Z basis of $\left| k \right \rangle_{A,Z}$, and Bob prepares his state in $\left| l \right \rangle_{B,Z}$. A prerequisite for a successful output is $k \neq l$. As each Z basis state propagates in different fiber, there is no crosstalk between them. According to the scheme of decoy state, Alice and Bob prepare their signal state in the weak coherent state $\left | \sqrt{\mu }e^{i\theta } \right \rangle$, here $\theta$ is totally randomized, $\mu$ is a parameter that characterizes the intensity of the pulse. In the Fock state representation, the probability $p_{n}$ for having the number state $\left| n \right \rangle$ is $e^{-\mu }\mu ^{n}/n!$. Considering the transmission rate $T=exp(-\alpha x)$ with $\alpha$ as the absorption coefficient and $x$ as the transmission distance, then the probability of obtaining a photon state from the Fock state $\left | n \right \rangle$ before hitting the single-photon detector is $T _{n}= 1-\left( 1-T \right)^{n}$. Due to the imperfection of the practical experiment setups, the background count rates from dark counts and stray light is $p_{dark}$, and the detection efficiency for the single photon detector is $\eta$ which is independent of the photon number. For the ease of discussion, we assume that both $p_{dark}$ and $\eta$ are equal for all single-photon detectors. So the probability for a single-photon detector to be fired when there is a photon number state $\left| m \right \rangle$ sent at the light source is
\begin{equation}
P_{m}=\eta T_{m}+p_{dark}-\eta T_{m}p_{dark}\approx \eta T_{m}+p_{dark}.
\end{equation}
\par Assume that the Fock state sent by Alice and Bob is $\left | m \right \rangle_{A}$ $\left | n \right \rangle_{B}$ with a gain of $Q_{mn}=p_{m}p_{n}$. In this case, there are mainly two kinds of detection events. The first kind of detection events are successful outputs in which two detectors are registered simultaneously in the position detection part. These successful outputs can be further divided into two sets, the right successful outputs and wrong successful outputs. For the right successful output, the indices of the registered detectors are $m$ and $n$, and at the same time there are no dark counts in other detectors. The probability for a right successful output is approximated as
\begin{equation}
P_{mn}=P_{m}P_{n}\left( 1-p_{dark} \right)^{N-2},
\end{equation}
where $N$ is the dimension of the basis. The second set of detection events are wrong successful outputs. An error may occur if one of the corresponding photon detector is not fired when a third detector fires due to dark counts. The probability for the fault detection of Alice's photon state is approximated as
\begin{equation}
P_{A,mn}=P_{B,n} \bar{P}_{A,m} p_{dark} (1-p_{dark})^{N-3},
\end{equation}
where $\bar{P}_{A,m}=1-P_{A,m}$. Likewise, the probability for the fault detection of Bob's photon state is
\begin{equation}
P_{B,mn}=P_{A,m} \bar{P}_{B,n}p_{dark} (1-p_{dark})^{N-3}.
\end{equation}
These wrong successful detection events only occur for the dimension $N \geqslant 3$, and there are total $\left( N-2 \right)$ fault detection events for each of them. In addition, there is another kind of wrong successful output, in which two detectors other than those corresponding the basis of the Fock state sent by Alice and Bob are fired. This detection event only occurs when the dimension $N\geqslant 4$. There are in total $A_{N-2}^{2}$ such detection events, each with a probability of
\begin{equation}
\bar{P}_{mm}=(1-P_{A,m})(1-P_{B,n})p_{dark}^{2}(1-p_{dark})^{N-4}.
\end{equation}
Because the probability for this detection event is so small, it is reasonable to set $\bar{P}_{mm}=0$ in the simulation.
\par The second kind of detection events are the fault outputs in which the number of detectors fired per time is not equal to 2. In these cases, the output results are always omitted, but this will cause finite information loss. The probability for these events is $P_{\gamma,mn}$
\begin{equation}
P_{\gamma , mn}=1-P_{mn}-\left ( N-2 \right) \left( P_{A,mn}+P_{B,mn} \right).
\end{equation}
\par Then the transition probability matrix Fock state $\left | m \right \rangle_{A}$ $\left | n \right \rangle_{B}$ is
\begin{equation}
P=
\begin{bmatrix}
P_{mn} & P_{A(B),mn} & \cdots & 0 & P_{\gamma, mn}\\
P_{A(B),mn} & P_{mn} & \cdots  & 0& P_{\gamma, mn}\\
 \vdots & \vdots  & \ddots   &\vdots &\vdots \\
0 & P_{A(B),mn} & \cdots  & P_{mn} & P_{\gamma, mn}
\end{bmatrix}.
\end{equation}
Here the transition probability matrix is a weakly symmetric channel with a rank equal to $A_{N}^{2}$.  The capacity for this channel is
\begin{equation}
\begin{split}
    R_{m,n}=\left(1-P_{\gamma, mn}\right)(\log A_{N}^{2}-H \left [ {P}'_{mn},{P}'_{A,mn},\cdots, {P}'_{B,mn}, \cdots \right ] ).
\end{split}
\end{equation}
Here ${P}'_{mn}$ and ${P}'_{A(B),mn}$ are the normalized probability with respect to $1-p_{\gamma,mn}$, which have the expressions
\begin{equation}
\begin{split}
&{P}'_{mn}=P_{mn}/\left( 1-P_{\gamma, mn} \right);\\
&{P}'_{A(B),mn}=P_{A(B),mn}/\left( 1-P_{\gamma, mn} \right).
\end{split}
\end{equation}
Equation (9) gives the maximum amount of information carried by Fock state $\left| m \right \rangle_{A} \left| n \right \rangle_{B}$ in the ideal case. However, from the consideration of security, only the quantum keys extracted from single -photon states are unconditional security, and the loss of information from multi-number states can probably be utilized by Eve. The processes of post processing like error correction and privacy amplification also cause information loss. So the maximum of secure information reads
\begin{equation}
\begin{split}
R=&\frac{A_{N}^{2}}{4N^{2}}\left ( 1-p_{\gamma,11 } \right ) Q_{1,1} ( \log A_{N}^{2}-H_{X} \left [ {P}'_{11},{P}'_{A,11},\cdots, {P}'_{A,11}, \cdots \right ] )\\
&-\frac{A_{N}^{2}}{4N^{2}}f\left ( e \right )\sum_{m,n}\left ( 1-p_{\gamma,mn} \right )Q_{m,n}H_{Z} [ {P}'_{mn},{P}'_{A,mn},\cdots, {P}'_{A,mn}, \cdots ].
\end{split}
\end{equation}
Here it should be noted that in equation (11) the coefficient of $1/4$ is the probability for the output of these two photon states from the beam splitter to position detection side or momentum detection side, and the coefficient of $\frac{A_{N}^{2}}{N^{2}}$ is the ratio for the instance that the basis of Alice's photon state differs from that of Bob. $H_{X} \left [ {P}'_{11},{P}'_{A,11},\cdots, {P}'_{A,11}, \cdots \right ]$ is the amount of information sacrificed for privacy amplification, and the coefficient $f\left( e \right)$ is the inefficiency of error correction, which always takes the value between 1.2 and 2 according to the error protocol. Here, we want to emphasize that there is a difference between the equation of secret key rate in our paper and that put forward by Lo et al.\cite{lo2012measurement}. In their work, the information loss is calculated by first evaluating the mean value of the error bit rate for all number states $\left| m \right \rangle_{A} \left| n \right \rangle_{B}$. While in our work, the information loss is evaluated for each of the number state $\left| m \right \rangle_{A} \left| n \right \rangle_{B}$. As the entropy is concave function of the probability distribution, so equation (8) in this paper increases the lower bound of the secret key rate.

\begin{figure}[ht]
\centering
\includegraphics[width=120mm]{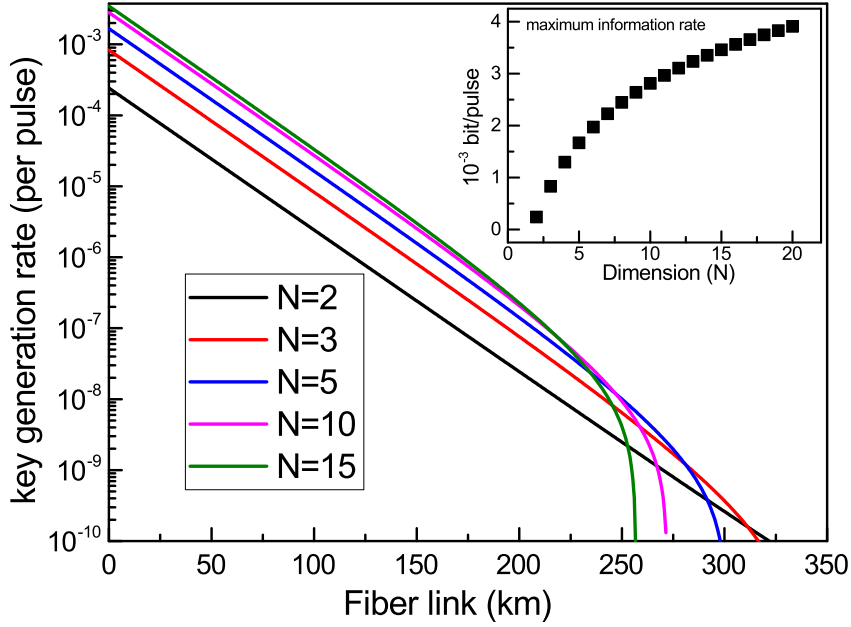}
\caption{Lower bound on the secret key rate for different dimensions of the high-dimensional MDI-QKD scheme given by equation (8) in logarithmic scale. In this simulation, we adopt the values of some parameters from ref \cite{lo2012measurement}: the loss coefficient of the fiber for 1550 nm is 0.2 dB/km, the background count rate is $6.02 \times 10^{-6}$, the detection efficiency of single-photon detector at 1550 nm is 20$\%$, and $\mu$ which determines the mean photon number per pulse of signal state is set to 0.5. Inset is the maximum secret key rate for each dimension at $x=0km$. In this simulation, we neglect the possible overlap in the diffraction peaks between adjacent momentum basis for low dimension, only focus on the dependence of secret key rate curve on the dimension $N$. In addition, to simplify the simulation, the lost entropy $H_{X} \left [ {P}'_{11},{P}'_{A,11},\cdots, {P}'_{A,11}, \cdots \right ]$ is substituted by $H_{Z} \left [ {P}'_{11},{P}'_{A,11},\cdots, {P}'_{A,11}, \cdots \right ]$.}
\label{Fig.}
\end{figure}

\par Here we evaluate the secret key rate for the realistic setup. The loss coefficient for the optical fiber at 1550 nm is $\beta=0.2 dB/km$ and $\alpha=(\beta \ln10)/10$, the detection efficiency at this wavelength is $14.5\%$ for a common commercial single photon detector, the dark count rate for these detectors is $6.02\times 10^{-6}$, the value of $\mu$ which characterizes the mean photon number per pulse of the signal state is set to $0.5$ according to ref \cite{lo2012measurement}, the value of the inefficiency function $f(e)$ is 1.2. Furthermore, we assume that all the fibers and detectors are identical. As is discussed above, to reduce the crosstalk between different measurement results for momentum measurement, the basis in momentum space is chosen sparsely. In this case, the amount of possible information leakage to Eve is much less than that in position space. To simplify the simulation, we use $H_{Z} \left [ {P}'_{11},{P}'_{A,11},\cdots, {P}'_{A,11}, \cdots \right ]$ to substitute $H_{X} \left [ {P}'_{11},{P}'_{A,11},\cdots, {P}'_{A,11}, \cdots \right ]$.

\par Fig. 3 plots the lower bound of secret key rate for different dimensions given by equation (8) in the high-dimensional MDI-QKD scheme. In this simulation, to focus on the dependence of the quantum secret key generation rate curve on the dimension $N$, we has neglected the possible overlap in the diffraction peaks between adjacent momentum basis for low dimension. From this figure, it can be seen that the secret key rate increases with the dimension $N$. The dependence of maximum secret key rate $R(x=0)$ on the dimension $N$ is shown in the inset. The secret key rate increases with the dimension, especially for small $N$. The increase rate is notable, as $N$ grows larger, the increase rate gradually slows down. The secret key rate $R$ shows an exponential behavior with respect to the transmission distance, but there is a critical point at which the secret key rate deviates from the exponential behavior and decreases more rapidly. This falling critical point determines the transmission distance of the high-dimension MDI-QKD. It can be seen in Fig.3 that the transmission distance decreases with the dimension $N$. Near the falling critical point, the probability of the wrong successful output is comparable to that of the right successful output, as $N$ grows larger, the proportion occupied by the wrong successful output becomes more significant, as well as the noise entropy. This causes a decrease in the transmission.

 \par In summary, we have proposed an experimental scheme to implement high-dimensional MDI-QKD by exploiting the discrete position basis and momentum basis as two MUBs. It has the same level of security as that of polarization-based two-dimensional MDI-QKD. Given a high enough dimension for this experimental proposal, not only the quantum secret key rate can be enlarged significantly, but also that the quantum-Nit error rate coming from the overlap between diffraction peaks of different basis states in X basis measurement will be reduced substantially. Furthermore, the transmission distance decreases as an increasing in dimension, but this drawback can be overcome by the method of quantum relay. A combination of high-dimensional MDI-QKD and existed optimization methods would serve as a rational approach for further improvement of secret key generation rate . 
 
 This work is supported by Young fund of Jiangsu Natural Science Foundation of China (SJ216025), National fund incubation project (NY217024), Scientific Research Foundation of Nanjing University of Posts and Telecommunications (NY215034), National Natural Science Foundations of China (Grant No. 61475075).
 
\bibliography{reference}

\end{document}